\documentclass[conference]{IEEEtran}
\IEEEoverridecommandlockouts

\usepackage{amsmath,amssymb,amsfonts}
\usepackage{algorithmic}
\usepackage{graphicx}
\usepackage{textcomp}
\usepackage{xcolor}
\usepackage{lipsum}
\usepackage{multirow}
\usepackage{multicol}
\usepackage{makecell}
\usepackage{url}
\def\BibTeX{{\rm B\kern-.05em{\sc i\kern-.025em b}\kern-.08em
    T\kern-.1667em\lower.7ex\hbox{E}\kern-.125emX}}
\usepackage[style=nature,sorting=none]{biblatex}
\usepackage[acronym]{glossaries}

\newacronym{EV}{EV}{Electric Vehicles}
\newacronym{V2G}{V2G}{Vehicle-to-Grid}
\newacronym{DER}{DER}{Distributed Energy Resources}
\newacronym{TSO}{TSO}{Transmission System Operator}
\newacronym{DSO}{DSO}{Distribution System Operator}
\newacronym{FCR}{FCR}{Frequency Containment Reserve}
\newacronym{aFRR}{aFRR}{automatic Frequency Restoration Reserve}
\newacronym{mFRR}{mFRR}{manual Frequency Restoration Reserve}
\newacronym{BSP}{BSP}{Balancing Service Providers}
\newacronym{BRP}{BRP}{Balancing Responsible Partie}
\newacronym{ICT}{ICT}{Information and Communication Techniques}

\makeglossaries

\begin{document}

\title{A review of potential barriers to the participation of electric vehicles in German balancing markets}

\author{\IEEEauthorblockN{Périne Cunat}
\IEEEauthorblockA{\textit{Master's student in Electrical Power Engineering} \\
\textit{RWTH Aachen}
}
}

\maketitle

\begin{abstract}
As conventional flexibility providers are gradually being replaced by variable renewable energies and electricity demand keeps rising, additional flexibility will become increasingly valuable for the power system.  Meanwhile, new sources of flexibility known as distributed energy resources are starting to emerge. Among them electric vehicles have a promising potential to provide short-term frequency-control services. However their integration into balancing markets faces various barriers. Based on the existing literature, this paper aims at identifying which potential barriers exist in the case of the participation of electric vehicles in German balancing markets.
\end{abstract}


\section{Introduction}
Supply-demand balance is one of the most important principles of power system operation and consists in maintaining a very fine balance between the energy drawn from the power system by the load and the energy injected into it by the generation facilities. This balance is closely linked to the frequency of the power grid, as an imbalance between demand and supply causes a deviation of the frequency from its target value.

The frequency of the power system is shared over large geographical areas called synchronous areas. On the European interconnected grid, for example, there are five synchronous areas where the nominal frequency has been set at 50Hz and match the nominal rotation speed of the synchronous generation units connected to the grid. In this case, a frequency higher than 50Hz indicates a surplus of generation compared to consumption and a frequency lower than 50Hz indicates a deficit of generation within the synchronous area.
It is essential that the frequency of the electrical network is kept within acceptable limits around its nominal value. Otherwise, the synchronous generation units could "lose their way" (loss of synchronism) and cascade off the grid, leading to a complete collapse of the power system in a few seconds (blackout). Fine control of the network frequency, i.e. the supply-demand balance, is therefore essential. \cite{SED2018}

\subsection{The role of balancing reserves for frequency control}
\begin{figure*}[t]
\centerline{\includegraphics[width=0.6\textwidth]{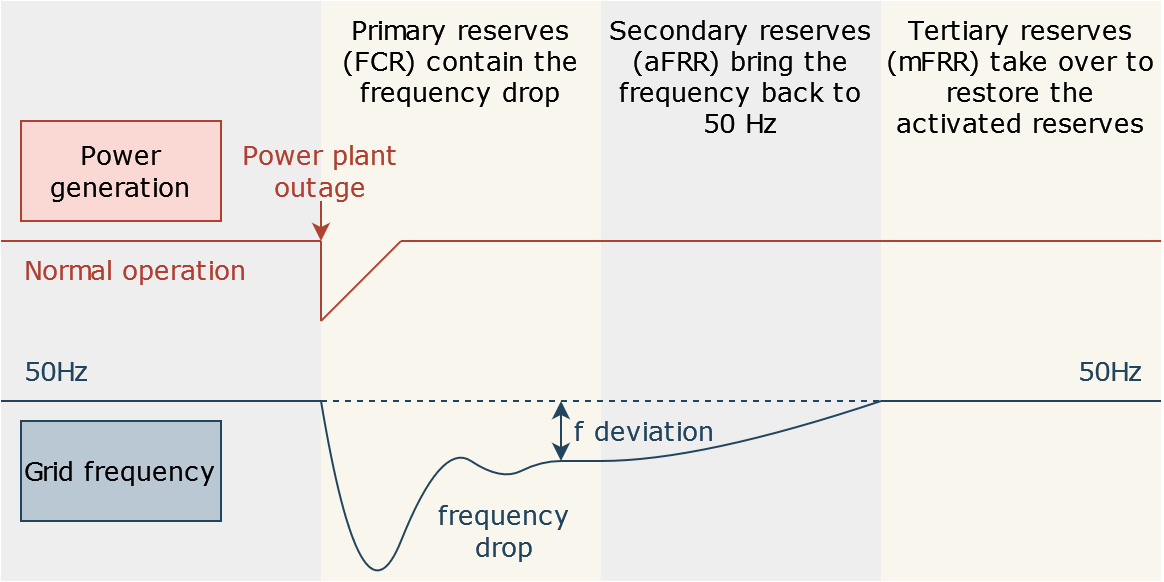}}
\caption{The sequential activation of power reserves (f: frequency)}
\label{fig:action_of_power_reserves}
\end{figure*}

The supply-demand balance is the responsibility of the Transmission System Operators (\acrshort{TSO}s) who monitor the power system to ensure real-time balance and maintain the frequency. For this purpose, consumption is continually assessed, from several years to real-time, to ensure that sufficient generation resources will be available to meet it and to plan the activation of each generation unit. 
However, the projected balance at a given moment is subject to uncertainties, which can come from production outages, errors in forecasting consumption and production of variable renewable energies... resulting in the frequency deviating from 50Hz. To compensate such deviations and maintain the balance in real time, the TSOs require dispatchable power reserves, both upward and downward. Several types of such power reserves can be distinguished, according to their speed and duration of activation:
\begin{itemize}
    \item Frequency Containment Reserves (\acrshort{FCR}) are almost instantaneously activated to stabilise frequency drops or spikes. FCR are the fastest type of reserves and are operated at the scale of a synchronous area in a joint process involving all TSOs. In continental Europe this reserve is tailored to compensate within 30 seconds to the loss of the two most important generators on the network, which represent 3GW of generation capacity,
    \item Frequency Restoration Reserves (FRR), which are either automatic (\acrshort{aFRR}) or manual (\acrshort{mFRR}), take over within a few minutes to restore the frequency back to its nominal value in a process involving only the TSO responsible of the area which is source of frequency deviation,
    \item Replacement Reserves that replace FCR and FRR to maintain long-term the restored frequency.
\end{itemize}
Figure \ref{fig:action_of_power_reserves} illustrates the respective action of each of these reserves in frequency control.

Today the majority of power reserves are provided by conventional power plants and can be purchased by TSOs through balancing markets. Such power plants provide flexibility to the power system: either positive flexibility by increasing their generation when the frequency is too low or negative flexibility by decreasing their generation when the frequency is too high.
\cite{RTE_EOD, DES_future, rte_reserves}

\subsection{Potential of electric vehicles to provide balancing power}
In the current context where conventional flexibility providers are gradually being replaced by variable renewable energies and developing electrification of many processes requires more electricity generation, additional flexibility will become an increasingly valuable resource to balance supply and demand in real time. Meanwhile, new sources of flexibility are starting to emerge, especially distributed energy resources (\acrshort{DER}s), which \cite{damsgaard_study_2015} defines as small- to medium- scale resources that are connected mainly to the lower voltage levels (distribution grids) of the system or near the end users. Among them, electric vehicles (\acrshort{EV}s) have technical characteristics that enable them to provide both positive and negative flexibility services:
\begin{itemize}
    \item either through smart charging, which consists in reducing the power consumed by EVs when the frequency is too low and increasing it when the frequency is too high (unidirectional power exchange),
    \item or in a more advanced way through vehicle-to-grid (V2G), which also uses the batteries of EVs as storage for the power system. Energy can either be injected from the batteries into the grid when the frequency is too low or stored from the grid into the batteries when the frequency is to high (bidirectional power exchange).
\end{itemize}
Due to their low energy capacity and high speed of activation, EVs are potentially best employed as a short-time power reserve. \cite{borne_barriers_2018}

The total storage capacity of electric vehicles in Germany is currently about 40GWh,  in comparison the capacity of Germany's largest pumped hydro power plant (Goldisthal) is 8,5GWh. \cite{figgener_development_nodate, Goldisthal_Vattenfall}. In the future, the storage capacity offered by electric vehicles should grow further with the expected electrification of the transport sector. This large potential of distributed storage is already integrated as an important flexibility lever in some long-term prospective scenarios for the development of the power system, such as those of the recent study published by RTE "Energy Pathways to 2050" \cite{FE2050}.

However, not all of this potential is always available for the power system as the first function of EVs remains transportation. EVs in travel can not be used to provide flexibility services and even though EVs are still most of the time, the power balancing capacity they can provide is constrained with charging requirements before the next ride and limited access to charging stations. Furthermore the available capacity of plug-in EVs depends on drivers' behaviour and is difficult to predict for a single vehicle. That is why the integration of EVs in balancing processes is most likely to happen through aggregation in EV fleet, whose characteristics are more statistically reliable and which allows to reach large enough capacity to participate in the balancing markets. Thus particular market players known as aggregators are expected to perform an important function of facilitating EVs', and more generally DERs', access to the balancing markets by acting as an intermediary between the owners of DERs and the markets.

\subsection{Aim of this paper}
Even though DERs and especially EVs have a great potential to provide balancing reserves and contribute to ensure security of supply, their integration into European balancing markets faces many barriers. These barriers can differ significantly from country to country.
Based on existing literature, the aim of this paper is to provide an insight of the features of the German balancing markets (Section \ref{section:GermanMarkets}) and propose a review of existing barriers to the integration of EVs in these markets, with focus on the provision on FCR and FRR (Section \ref{section:Barriers}).

\section{Literature review}

\begin{table*}[t]
\centering
\caption{Barriers to the integration of DERs into the balancing markets identified in the reviewed literature\\(* articles specifically dealing with electric vehicles)}
\label{table:LitteratureSynthesis}
\begin{tabular}{ |c|c|c|c|c|c|c|c|c| } 
 \hline
 Type & \makecell{Barriers to the integration \\of DERs into balancing markets} & \cite{codani_missing_2014}* & \cite{borne_barriers_2018} & \cite{poplavskaya_distributed_2019} & \cite{barbero_critical_2020} & \cite{plaum_aggregated_2022} & \cite{lauinger_review_nodate}* & \cite{sovacool_future_2017}* \\
 \hline
 \multirow{3}{*}{Regulatory} & Explicit discrimination against DERs & & x & x & x & x & & \\
 & Prequalification requirements & & & x & & x & & \\
 & Additional agreements & & & x & x & x & & \\
 \hline
 \multirow{2}{*}{Market access} & Minimum bid size & x & x & x & x & x & & \\
 & Symmetry of products & x & x & x & x & x & & \\
 & Time-related requirements & & x & x & x & x & & \\
 \hline
 \multirow{4}{*}{Economic} & High investment costs & & & & x & x & & x \\
 & High penalization costs & & & & x & x & & \\
 & Nature of the remuneration & x & x & x & x & & & \\
 & Inadequate subsidization & & & x & x & x & & \\
 \hline
 \multirow{2}{*}{Technical} & Lacking ICT & & & & & x & & x\\
 & Interoperability between TSOs & x & x & & & x & & \\
 \hline
\multirow{1}{*}{Behavioural} & Range anxiety & & & & & & x & x\\
 \hline
\end{tabular}
\end{table*}

The use of EVs to provide flexibility services is the subject of much research today.

In 2014 \cite{codani_missing_2014} proposed a first attempt to identify potential barriers to the integration of EVs in balancing markets by defining a modular framework to analyse the market designs for reserves. This framework has two modules. The first one includes the "rules toward the aggregation of EVs", such as technical discrimination against aggregated resources and interoperability among \acrshort{DSO}s. The second module contains the "rules defining the payment scheme of grid services", such as the nature of payment and extra remuneration bonus for flexibility.

This work was deepened and extended to all DERs in 2018 by \cite{borne_barriers_2018} who added a module containing "rules to define the products on the markets", hierarchised the three resulting modules and applied the new framework to compare balancing markets in four European countries including Germany. \cite{poplavskaya_distributed_2019} and \cite{barbero_critical_2020} continued the work of \cite{borne_barriers_2018} and proposed frameworks to classify the barriers for DER aggregators.

From the framework developed in \cite{poplavskaya_distributed_2019}, the barriers for aggregators to enter the balancing markets can largely be classified as either market access oriented or auction configuration oriented, where market access oriented barriers include formal access requirements, administrative aspects, technical prequalification criteria, while auction configuration oriented barriers include bid-related requirements, time-related characteristics, and remuneration. The framework developed in \cite{barbero_critical_2020} classifies barriers as either regulatory, technical or economic and was applied in a study case of German balancing markets.

Barriers identified in \cite{borne_barriers_2018}, \cite{poplavskaya_distributed_2019}, \cite{barbero_critical_2020} were reviewed in \cite{plaum_aggregated_2022}. In addition to them, the reviews proposed by \cite{lauinger_review_nodate} and \cite{sovacool_future_2017} in 2017, also identified behavioural barriers, such as range anxiety which refers to the concern of EVs owners that participating in balancing processes would reduce the distance their vehicle can travel between charges.

Table \ref{table:LitteratureSynthesis} summarises the barriers identified in the literature that will be discussed in Section \ref{section:Barriers} and the articles in which they appear.

\section{Fundamentals of balancing markets in Germany}
\label{section:GermanMarkets}

\begin{table*}[t]
\centering
\caption{Auction configuration for the procurement of primary and secondary reserves in Germany}
\label{table:AuctionConfiguration}
\begin{tabular}{ |c|c c c | } 
 \hline
 Bid related requirements & FCR & aFRR & mFRR \\
 \hline
 Minimum bid size & 1MW & 1MW & 1MW \\ 
 Bid symmetry & symmetrical & asymmetric & asymmetric \\
 \hline \hline
 Time related requirements & FCR & aFRR & mFRR \\
 \hline
 Frequency of bidding & daily & daily & daily\\
 Frequency of market clearing & once a day &  \makecell{once a day (capacity) \\once per 15 min (energy)} & \makecell{once a day (capacity) \\once per 15 min (energy)}\\
 Product resolution & 4 hours & 4 hours & 4 hours \\
 \hline
\end{tabular}
\end{table*}

This section provides an insight of most important features of German balancing markets and introduces all necessary notions to the discussions lead in section \ref{section:Barriers}.

\subsection{Actors of the balancing markets}
The German balancing markets are operated by four TSOs (50Hertz, Amprion, TenneT, TransnetBW) who act as the single buyers in these markets. Close cooperation between German TSOs has been implemented by the introduction of the Grid Control Cooperation, which allows Germany to be considered as a uniform market for flexible products, with common rules and coordinated actions between TSOs through a common platform (\url{www.regelleistung.net}).

On the supply side, market participants with reserve-providing groups known as Balancing Service Providers (\acrshort{BSP}s) offer balancing services by formulating capacity and energy bids for each balancing products. The amount of balancing needed is determined through the balance reserve sizing process. The bids are then activated according to the merit order, i.e. the cheapest bids are activated first, when imbalances in demand and supply occur.

In Germany, each supplier or consumer must be part of a so-called Balancing Responsible Partie (\acrshort{BRP}), either directly or through an intermediary such as a supply company. BRPs are responsible for the balance of their portfolios, so that TSOs only compensate for the remaining imbalances that occur when the energy volume physically injected into/withdrawn from the power system by the BRPs differs from the one declared beforehand to the TSOs at intraday gate closure. \cite{poplavskaya_distributed_2019, DES_future, zeller_main_nodate}

As mentioned in the introduction, aggregators are gaining in importance in balancing markets by allowing the participation of DERs. \cite{poplavskaya_not_2018} defines these emerging markets players as "market participants that combines multiples customer loads or generated electricity for sale, for purchase or auction in any organised energy market". According to \cite{poplavskaya_not_2018}, in 2018 eight independent aggregators had been prequalified to participate in the balancing markets in Germany for one or several flexibility products with portfolios including a variety of DERs, such as Combined Heat and Power units, industrial loads and power-to-heat, as well as renewable generation and energy storage.

\subsection{Flexibility products}

The following flexibility products are currently implemented in Germany:
\begin{itemize}
    \item Primary reserve (FCR)
    \item Secondary reserve (aFRR)
    \item Tertiary reserve (mFRR)
    \item Quickly and immediately interruptible loads (restricted to large consumption sites such as industrial customers)
\end{itemize}

Table \ref{table:AuctionConfiguration} contains the main parameters of the auction configuration for the provision of FCR or FRR and refers to the latest flexibility products definition delivered in 2020. \cite{FCRregelleistung, aFRRregelleistung, mFRRregelleistung} Here are some indications for a good understanding of this table. A symmetrically specified product, is a product for which balancing assets must be able to provide both negative and positive flexibility in same height. The frequency of bidding determines how often bids are called: capacity bids for FCR and jointly formulated capacity and energy bids for FRR. The frequency of market clearing determines how often a merit order of capacity bids is built. The product resolution refers to the timeframe of subproducts traded within the same bidding period.

\begin{figure}[h]
\centerline{\includegraphics[width=0.5\textwidth]{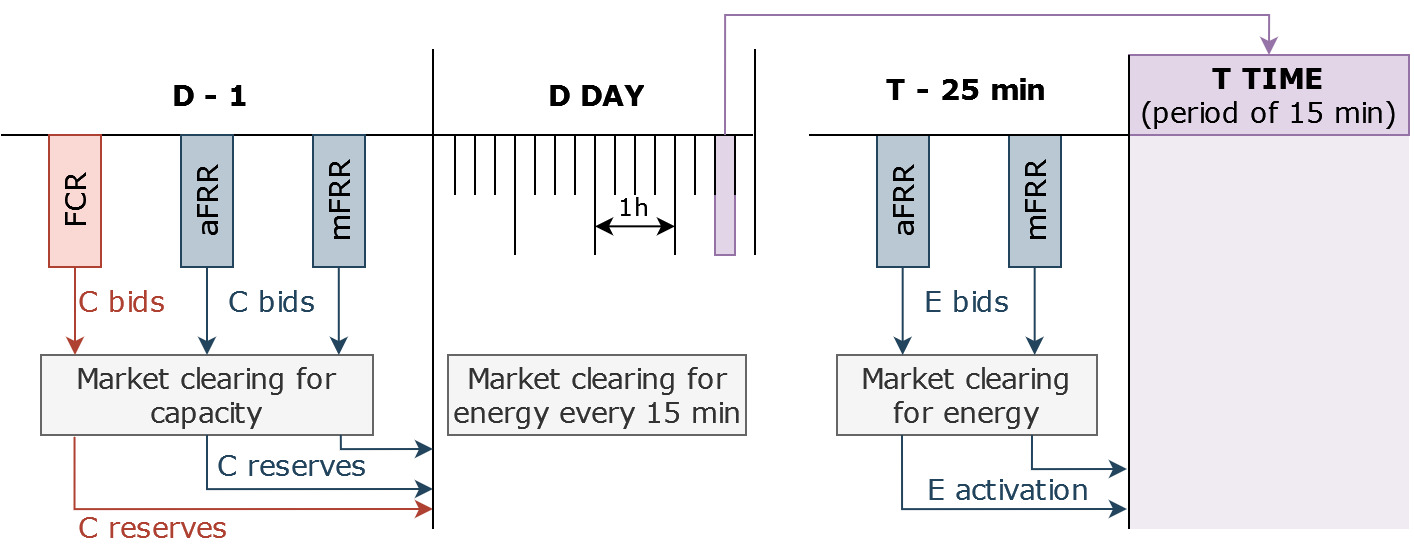}}
\caption{Timing for the procurement of FCR, aFRR and mFRR products in German balancing markets (C: Capacity, E: Energy, min: minutes)}
\label{fig:GermanBalancingMarkets}
\end{figure}

For German FCR, the procurement is conducted on a daily basis with calls for bids taking place the day prior each contract's start. Bids for FCR are related to a 4 hours period and are selected according to a financial merit order based on their submitted price (€/MW of available capacity). As for FRR, the formulation of bids are also made daily for both automatic and manual reserves. The submitted offers are selected based on the merit order for availability price (€/MW of available capacity), and are then activated in the merit order for utilisation price (€/MWh exchanged with the power system) taking place every 15 minutes within the contract's period. This is referred to as differentiated activation. Figure \ref{fig:GermanBalancingMarkets} illustrates these processes over time. \cite{poplavskaya_distributed_2019, zeller_main_nodate}


\subsection{Type of remuneration}
We can distinguish between two sources of remuneration for flexibility products:
\begin{itemize}
    \item An availability remuneration for reserving capacity to provide frequency-control service if it is needed,
    \item An utilisation remuneration for the amount of energy provided to the power system if balancing capacity was activated.
\end{itemize}
In the case of German FCR, only availability remuneration is considered whereas availability and utilisation remunerations are applied for the provision of FRR.


\section{Potential barriers to the integration of electric vehicles in German balancing markets}
\label{section:Barriers}
As developed in the introduction, electric vehicles represent a new potential source of flexibility for the power system. However, some obstacles remain regarding their integration in the balancing markets. In this section we try to identify what these barriers might be in the case of the German markets.
For this purpose we will follow the classification proposed by \cite{plaum_aggregated_2022} by considering regulatory, market access related, economic and technical barriers, to which we add behavioural barriers identified in \cite{lauinger_review_nodate} and \cite{sovacool_future_2017}.

\subsection{Regulatory barriers}
In Germany there is no explicit discrimination against DERs in the regulatory frameworks of balancing markets : unrestricted pooling is allowed for aggregators (no constraints regarding the technologies used in the portfolio or the number of units) and BSPs can be located at any voltage levels of the power system. \cite{zeller_main_nodate}
However administrative aspects such as additional agreements with a BRP or prequalification criteria may prevent EVs from being allowed to participate in balancing markets.

\subsubsection{Additional agreements} they are a specific concern for independent aggregators, which are defined by \cite{poplavskaya_not_2018} as aggregators which are not affiliated to a supplier or any other market participant. In Germany independent aggregators must coordinate with a BRP to deal with the imbalances that their portfolios may produce. Independent aggregators’ activities cause thus a higher administrative effort for their BRP due to schedule adjustments and exchanges as well as higher risks for the BRPs of those customers whose units are used for the provision of balancing energy. This results in additional financial charges for the aggregator, which is deterrent to their participation in balancing markets. \cite{poplavskaya_distributed_2019}

\subsubsection{Prequalification criteria} they include activation speed and duration, as well as ramp rate requirements. Because plug-in EVs can react very quickly to a control signal, they can comply with activation speed and ramp rate requirements. \cite{codani_missing_2014} However activation duration can be more difficult to meet for EVs. Indeed their flexibility potential depends on daily usage patterns, which limits the amount of stable power output a EV fleet can provide throughout an extended activation period. \cite{poplavskaya_distributed_2019}

Following the shortening of German flexibility products' resolution in 2020 from a weekly basis to a few hours, the minimum activation duration requirements are today less restrictive. According to the revised prequalification criteria of June 2022 \cite{prequalification_requirements} the marketable power of a flexibility providing asset is defined as "the power that can be continuously and fully provided over a period of four hours (FRR) or two hours (FCR)". EVs can more easily comply with these new criteria, which reduces the importance of the barrier that the prequalification criteria can represent.

\subsection{Market aspects related barriers}
The rules defining the products of the balancing markets have been created considering the characteristics of conventional generation units that have covered the flexibility needs of the power system until today. For assets with different characteristics, such as EVs, these rules may constitute barriers to formulate bids in balancing markets.

\subsubsection{High minimum bid size}
Minimum bidding amount to be allowed to participate in balancing markets is a rule cited in almost all articles of this paper's literature review as a barrier to the integration of DERs in balancing markets, \cite{borne_barriers_2018} and \cite{poplavskaya_distributed_2019} identify it clearly in their study cases of German markets. Even though, the minimum bids for providing balancing services in Germany were decreased in 2020, it still requires aggregators to have a large pool of small-scale providers to comply.


In the case of EVs, we can convert the minimum bidding-amount into a minimum number of EVs, as proposed by \cite{codani_missing_2014}. To measure the evolution of this barrier, we make the calculation for the years 2014 and 2021. We then compare the required number of EVs to the German national fleet of the corresponding year, based on the data of \cite{statista_nb_electric_cars_G}. If we consider charging stations of 7kW (domestic plugs) and that on average one EV out of three is available for reserve markets (taking into account that only a part of the fleet would be available to provide flexibility services due to transportation, charging needs, limited access to charging stations...), we obtain the results of Table \ref{table:MinimumNbEVs}.

With this simplified calculation, the minimum amount of EVs to be aggregated represented 17\% of the national fleet in 2014, at which time \cite{codani_missing_2014} identified minimum size as a significant barrier to the integration of EVs into balancing markets. Since then, thanks to the strong increase in the number of EVs in Germany and the decrease in the minimum bidding amount, minimum bid size is less of a barrier today, as the required amount of EVs is about 0.15\% of the national fleet under our assumptions.

\begin{table}[h]
\centering
\caption{Minimum amount of EVs to participate in German balancing markets in 2014 and 2021}
\label{table:MinimumNbEVs}
\begin{tabular}{ |c|c c c | } 
 \hline
  & \makecell{German \\national fleet} & \makecell{Minimum \\ bid size} & Required amount of EVs \\
 \hline
 2014 & 12,150 & 5MW & \makecell{2145\\ (17.5\% of the national fleet)}\\
 \hline
 2021 & 309,000 & 1MW  & \makecell{430\\ (0.15\% of the national fleet)}\\
 \hline
 \end{tabular}
 \end{table}

\subsubsection{Symmetry of flexibility products}
is also a rule often cited in the considered literature. When applied, as it is the case for German FCRs, it prevents DERs that can only provide one-way flexibility from participating in balancing markets. This is the case for DERs that provide flexibility services by reducing their consumption according to the needs of the system.

In the case of EVs, adapted control strategies could ensure the provision of both positive and negative flexibility. For V2G, there are for instance control strategies that keep the EV batteries in a specific range of states of charge to ensure both positive and negative flexibility, on the same principle as the method described by \cite{hollinger_distributed_2016} for photovoltaic module battery operation.

\subsubsection{Time-related requirements} As well as activation duration prequalification criteria, large product resolution can prevent EVs from participating in the provision of balancing reserves due to the difficulty of ensuring that the required power output will remain available during the entire duration of a bid. The shortening of German flexibility products' resolution in 2021 to 4 hours has eased the constraints for formulating a bid and enabling EVs to participate in balancing markets.

\subsection{Economic barriers}
As the bids accepted by the TSOs to provide the required balancing power are market-based, EVs will only be able to participate in the balancing markets if they are economically profitable enough. Besides high investment costs to deploy the required \acrshort{ICT} infrastructures, additional costs due to faster aging of the batteries of flexibility providing EVs and also penalization costs in case the aggregators' portfolios are not able to provide the amount of balancing power formulated in the bids (due to availability uncertainties), some of the literature considers that the remuneration and subsidization methods in Germany are currently not the most favourable to promote EVs as flexibility providers.

\subsubsection{Nature of the remuneration} as argued by \cite{codani_missing_2014}, remuneration which is not only composed of availability remuneration but also utilisation remuneration are more attractive for fast-ramping resources such as EVs for they exchange more MWh with the grid than slow-ramping units at equal capacity. Under these considerations, the remuneration of German FCR is not optimal for the economic efficiency of EVs in balancing markets.

\subsubsection{Inadequate subsidization} in Germany, flexibility subsidies are granted to biogas plants which encourage their wider use for balancing. Other types of DERs and especially EVs do not benefit from such subsidies, which can be deterrent to their economic efficiency and thus to their entry into balancing markets.  \cite{poplavskaya_distributed_2019}

\begin{table*}[t]
\centering
\caption{Synthesis of identified barriers to the integration of electric vehicles in current German balancing markets}
\label{table:BarriersSynthesis}
\parbox{.56\linewidth}{
\begin{tabular}{ |c|c|c|c|c| } 
 \hline
 Type & \makecell{Discussed barriers} & FCR & aFRR & mFRR \\
 \hline
 \multirow{3}{*}{Regulatory} & Explicit discrimination against DERs & - - & - - & - - \\
 & Prequalification requirements & - & - & - \\
 & Additional agreements & + & + & + \\
 \hline
 \multirow{2}{*}{Market access} & Minimum bid size & - & - & - \\
 & Symmetry of products & - - & - - & - - \\
 & Time-related requirements & - & - & -\\
 \hline
 \multirow{4}{*}{Economic} & High investment costs & + & + & + \\
 & High penalization costs & + & + & + \\
 & Nature of the remuneration & + & - - & - - \\
 & Inadequate subsidization & + & + & + \\
 \hline
 \multirow{2}{*}{Technical} & Lacking ICT & + & + & +\\
 & Interoperability between TSOs & / & / & /\\
 \hline
\multirow{1}{*}{Behavioural} & Range anxiety & / & / & /\\
 \hline
\end{tabular}
}
\parbox{.19\linewidth}{
\begin{tabular}{|c|c|}
\hline
- - & no barrier\\
-  & decreasing barrier\\
+  & steady barrier\\
/  & lacking information\\
\hline
\end{tabular}
}
\end{table*}

\subsection{Technical barriers}
Even if aggregated EVs met all the conditions for accessing balancing markets and were economically efficient enough to participate, their actual implementation depends on technical prerequisites. The first one would be the deployment of suitable ICT infrastructure to ensure the collection of availability data besides monitoring and controlling of EVs by the aggregator. Another one is interoperability among DSOs, which is of great importance for the EVs of an aggregator's portfolio could use charging stations located in different DSOs' operation areas and move from one to another within one day, especially in Germany where over 800 different DSOs operate. \cite{codani_missing_2014} Although this point has been raised in several articles, little information seems to be available concerning the current level of interoperability between German DSOs.

\subsection{Behavioural barriers}
Besides the barriers discussed so far in this paper, customer-based research has identified potential behavioural barriers to the participation of EVs in balancing markets, that should not be overlooked as it is the owners of EVs who will ultimately decide whether their vehicle will participate in frequency control. In this paper we will only discuss range anxiety, which refers to the concern of EVs owners that participating in balancing processes would reduce the distance their vehicle can travel between charges.

Range anxiety is fed  by the potential inconvenience of an EV's participation in flexibility services, if the battery state-of-charge control strategy is not sufficiently effective in meeting the transportation needs. Both smart charging and V2G could indeed lead to delayed or slower speed of charging and insufficient state of charge in case of urgent need of the EV. \cite{sovacool_future_2017}

In 2021, \cite{baumgartner_persisting_nodate} conducted a survey in Germany to investigate which range requirements in the context of V2G exist. The results showed relatively high range requirements, especially when compared to actual travelled distances and concluded that range anxiety could be a major barrier to the implementation of V2G. However the authors call for further research on the subject to complete this survey.

\section{Conclusion}

The power system is being transformed today under the impulse of the energy transition, which implies finding new sources of flexibility to continue to guarantee the security of supply. EVs show good potential to help cover future flexibility needs but several barriers remain to their integration into German balancing markets. However increasing number of EVs and recent revision of the definition of German flexibility products in 2020 (lower minimum bid size, higher bidding frequency) tend to ease some of these obstacles, as discussed in Section \ref{section:Barriers}. Table \ref{table:BarriersSynthesis} summarizes the results of this paper and indicates for each reviewed barrier if it:
\begin{itemize}
    \item doesn't currently exist in Germany (no barrier),
    \item has been potentially reduced compared to what was reported in the literature (decreasing barrier),
    \item is still an important barrier, as reported in the literature (steady barrier).
\end{itemize}

\section{Acknowledgement}
I would like to acknowledge and express my appreciation for the feedback provided by my supervisor, Claire Lambriex, during the development of this article. Her insights and suggestions have been valuable in refining my work.




\medskip

\printglossaries

\printbibliography

\end{document}